\documentclass[final, 5pt, authoryear, times]{article}
\newcommand{\myauthor}{A. Martin}

\newcommand{\mytitle}{Eshelby-based homogenization schemes with finite circular cylinders}
\title{\mytitle}
\author{\myauthor}

\usepackage{parskip}
\usepackage[fleqn]{amsmath}
\usepackage[top=3cm,bottom=4cm,marginparwidth=2cm]{geometry}

\usepackage{amssymb}
\usepackage{amsthm}
\newtheorem{remark}{Remark}

\usepackage{bm}
\usepackage{tikz}

\usepackage{xcolor}

\newcommand{\D}{\mathrm d}
\newcommand{\dbldot}{\mathbin{\mathord{:}}}

\newcommand{\eff}{\mathrm{eff}}

\newcommand{\reals}{\mathbb R}

\newcommand{\tens}[1]{\vec{#1}}

\renewcommand{\vec}[1]{\bm{\mathrm{#1}}}

\begin{document}
\maketitle

Commonly, for homogenization of fibrous media, fibers are approximated by ellipsoidal inclusions. Indeed, the solution of Eshelby's problem for an ellipsoid is well-known analytically. However, for a cylinder, the analytical solution is not easy to compute, and the internal field is not uniform (which makes the Hill tensor useless).
We here propose to give some tools for computing main homogenization schemes based on Eshelby's problem, for finite circular cylinders.
This document is also a companion to \cite{Martin2024}, where homogenization schemes like Dilute Scheme, Mori-Tanaka scheme \cite{Mori} and Ponte Casta\~neda \& Willis scheme \cite{Ponte} are used.

\section{Introduction and notations}
\begin{figure}
\begin{center}
\begin{tikzpicture}
\tikzstyle{operation}=[->,>=latex]
\fill[gray!20,rotate=-30] (-0.4,-2) -- (-0.4,2) -- (0.4,2) -- (0.4,-2) ;
\foreach \t in {-2,2}{
\filldraw[fill=gray!20,line width=0.7pt,rotate=-30] plot[domain=0:2*pi]
({0.4*cos(\x r)}, {\t+0.1*sin(\x r)});
}
\draw[line width=0.5pt,style=dashed,rotate=-30] plot[domain=0:2*pi]
({0.4*cos(\x r)}, {0.1*sin(\x r)});
\draw[line width=0.7pt,rotate=-30] (-0.4,-2) -- (-0.4,2) ;
\draw[line width=0.7pt,rotate=-30] (0.4,-2) -- (0.4,2) ;
\draw (-0.5,0.7) node {$\vec{t}_{\alpha}$} ;
\draw (1.1,-0.1) node {$\vec{s}_{\alpha}$} ;
\draw (0.5,1) node {$\vec{n}_{\alpha}$} ;
\draw[operation,rotate=-30] (0,0) to (0,1) ;
\draw[operation,rotate=-30] (0,0) to (0.9,0.2) ;
\draw[operation,rotate=-30] (0,0) to (-0.6,0.2) ;
\end{tikzpicture}
\end{center}
\caption{The cylinder basis}
\label{schemavar}
\end{figure}
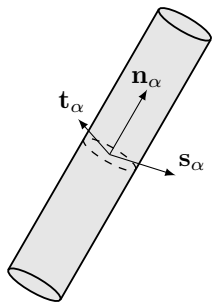

We consider a circular cylinder $\alpha$ whose radius is $R$ and length is $2L$, and introduce its aspect ratio $e=L/R$, and its isotropic stiffness $\tens C_\alpha$. Its Young modulus is noted $E_\alpha$, its shear modulus $\mu_\alpha$, and its Poisson's coefficient $\nu_\alpha$.
We consider a global orthonormal basis $(\vec e_1,\vec e_2, \vec e_3)$.
Cylinder $\alpha$ is oriented by the unit-vector $\vec n_\alpha$. This unit vector is parametrized by $(\theta,\phi)$ so that its components in the global basis are $(\sin\theta\cos\phi,\sin\theta\sin\phi,\cos\theta)$.
We introduce an orthonormal basis $(\vec s_\alpha, \vec t_\alpha, \vec n_\alpha)$ more suited to the cylinder, such that the components of $\vec s_\alpha$ in the global basis are $(\cos\theta\cos\phi,\cos\theta\sin\phi,-\sin\theta)$ (see Fig.~\ref{schemavar}).

Coordinates of tensors in the global basis $(\vec e_1,\vec e_2, \vec e_3)$ will be indexed by $1,2,3$ (and lowercase letters $i,j,k...$ in abstract notation), whereas coordinates in the cylinder basis $(\vec s_\alpha, \vec t_\alpha, \vec n_\alpha)$ will be indexed by $s,t,n$ (and $I,J,K...$ in an abstract notation)

\begin{remark}
Considering a given cylinder, the orthonormal basis $(\vec s_\alpha, \vec t_\alpha, \vec n_\alpha)$ is fixed. It should not be confused with the cylindrical coordinate basis which turns around $\vec n_\alpha$ (noted in \cite{Martin2024} $(\vec e_{r,\alpha},\vec e_\theta, \vec n_\alpha)$).
\end{remark}

\begin{remark}
The orthonormal basis $(\vec s_\alpha, \vec t_\alpha, \vec n_\alpha)$ will also be used in the following considering prolate ellipsoids (for which smallest semi-axes have the same length). The expression 'fiber basis' will be used. The aspect ratio $e$ will refer to the aspect ratio of the prolate ellipsoid. The stiffness of the ellipsoid will also be noted $\tens C_\alpha$.
\end{remark}

We note $\tens J$ (resp. $\tens K$)
the spherical (resp. deviatoric) fourth-rank projection tensor. More
precisely, $\tens J=\dfrac 13 \tens 1\otimes\tens 1$ and $\tens K=\tens I-\tens J$, where $\tens 1$ (resp. $\tens I$) is the second-
(resp. fourth-) rank identity tensor. We then have in an abstract notation:
\begin{equation}
\tens 1=\delta_{ij}, \quad \tens I=\frac12\left(\delta_{ik}\delta_{jl}+\delta_{il}\delta_{jk}\right), \quad \tens J=\frac13 \delta_{ij}\delta_{kl}.
\end{equation}
We will also use Voigt notation for fourth-rank tensor (see Appendix for details on this notation).

\section{Well-known formulas for the strain concentration tensor of an ellipsoid}
We consider a prolate ellipsoid $\alpha$ embedded in a homogeneous matrix whose isotropic stiffness is $\tens C_0$ (and shear modulus $\mu_0$, Poisson's coefficient $\nu_0$ and Young modulus $E_0$),
submitted to a uniform strain field $\tens{\overline{E}}$ at infinity. The ellipsoid has two identical small semi-axes whose common length is $a$ and a large semi-axis whose length is $b$, so that the ellipsoid aspect ratio is $e=b/a$.
The strain field $\varepsilon_\alpha$ on the ellipsoid is uniform and given by
\begin{equation}
\label{ellipsoid}
\varepsilon_\alpha=\tens A(\vec n_\alpha)\dbldot\tens{\overline{E}}
\end{equation}
where $\tens A(\vec n_\alpha)$, named \emph{concentration tensor}, is a fourth-rank tensor which depends on the normal vector $\vec n_\alpha$.
For ellipsoidal inclusions, the formulas for the concentration tensor are already known:
\begin{equation}
\label{A_ellipso}
\tens A(\vec n_\alpha)=\left[\tens I+\tens S_0(\vec n_\alpha)\dbldot\tens C_0^{-1}\dbldot\left(\tens C_\alpha-\tens C_0\right)\right]^{-1}
\end{equation}
where $\tens S_0$ is the Eshelby tensor, and its coordinates can be found for example in \cite{torq2002}.

If we first assume that contrast $\chi=E_\alpha/E_0$ is infinite, we have:
\begin{equation}
\tens A(\vec n_\alpha)=\tens C_\alpha^{-1}\dbldot\tens C_0\dbldot\left[\tens S_0(\vec n_\alpha)\right]^{-1}.
\end{equation}
And if our prolate ellipsoid has a high aspect ratio, we find the following first order approximations for the coordinates of $\tens S_0(\vec n_\alpha)$:
\begin{equation}
\begin{aligned}
&S_{nnnn}=\frac{4-2\nu_0}{2(1-\nu_0)}\frac{\ln e}{e^2},\qquad S_{nnss}=S_{nntt}=-\frac{1-2\nu_0}{2(1-\nu_0)}\frac{\ln e}{e^2}\\
&S_{ssss}=S_{tttt}=\frac{5-4\nu_0}{8(1-\nu_0)},\qquad S_{ssnn}=S_{ttnn}=\frac{\nu_0}{2(1-\nu_0)}\\
&S_{sstt}=S_{ttss}=\frac{\nu_0}{2(1-\nu_0)},\qquad S_{stst}=\frac{3-4\nu_0}{8(1-\nu_0)},\qquad S_{snsn}=S_{tntn}=\frac 14,
\end{aligned}
\end{equation}
and the other components are null (see \cite{torq2002}). These coordinates can be written in Voigt notation (if we consider the basis $(\vec s_\alpha, \vec t_\alpha, \vec n_\alpha)$):
\begin{equation}
\tens S_0(\vec n_\alpha)=
\begin{pmatrix}
S_{ssss}&S_{sstt}&S_{ssnn}&0&0&0\\
S_{sstt}&S_{ssss}&S_{ssnn}&0&0&0\\
S_{nnss}&S_{nnss}&S_{nnnn}&0&0&0\\
0&0&0&1/2&0&0\\
0&0&0&0&1/2&0\\
0&0&0&0&0&2S_{stst}\\
\end{pmatrix}
\end{equation}
We then obtain the coordinates of $\tens A(\vec n_\alpha)$ in the fiber basis (in Voigt notation):
\begin{equation}
\tens A(\vec n_\alpha)=
\begin{pmatrix}
&&&&&\\
&\tens A_{1}&&&(0)&\\
&&&&&\\
&&&\frac{2\mu_0}{\mu_\alpha}&0&0\\
&(0)&&0&\frac{2\mu_0}{\mu_\alpha}&0\\
&&&0&0&\frac{\mu_0}{2S_{stst}\mu_\alpha}\\
\end{pmatrix},\quad\text{with  }
\tens A_1=\tens C_1^{-1}\cdot\tens C_{2}\cdot\tens S^{-1}
\end{equation}
where
\begin{equation}
\tens C_1=
\begin{pmatrix}
C^\alpha_{11}&C^\alpha_{12}&C^\alpha_{13}\\
C^\alpha_{21}&C^\alpha_{22}&C^\alpha_{23}\\
C^\alpha_{31}&C^\alpha_{32}&C^\alpha_{33}\\
\end{pmatrix},\quad \tens C_{2}=
\begin{pmatrix}
C^0_{11}&C^0_{12}&C^0_{13}\\
C^0_{21}&C^0_{22}&C^0_{23}\\
C^0_{31}&C^0_{32}&C^0_{33}\\
\end{pmatrix},\quad \tens S=
\begin{pmatrix}
S_{ssss}&S_{sstt}&S_{ssnn}\\
S_{sstt}&S_{ssss}&S_{ssnn}\\
S_{nnss}&S_{nnss}&S_{nnnn}\\
\end{pmatrix}
\end{equation}
Computation of $\tens A_1$ easily shows that all components of $\tens A(\vec n_\alpha)$ are proportional to $1/\chi$, which is small.
Moreover, for high aspect ratios, $A_{nnnn}$ and $A_{ssnn}$ are proportional
to $e^2/(\chi\ln e)$, and therefore more significant than all other components. This will be important in the following.

\begin{remark}
If we assume first the aspect ratio infinite, and then the contrast infinite,
the expressions are not the same, but we can show that the most significant terms
remain $A_{nnnn}$ and $A_{ssnn}$.
\end{remark}

\section{Closed-form formulas for the strain concentration tensor of a finite circular cylinder}
\label{sec_closed}
Here, we consider the case of a finite circular cylinder $\alpha$, embedded in the same matrix
as above, submitted to the same uniform field $\tens{\overline{E}}$ at infinity.
The strain field is no more uniform on the cylinder, but we can consider its
average on the fiber, that we note $\varepsilon_\alpha$. Again, we can note
\begin{equation}
\label{cylinder}
\varepsilon_\alpha=\tens A(\vec n_\alpha)\dbldot\tens{\overline{E}}
\end{equation}
where $\tens A(\vec n_\alpha)$ is a fourth-rank tensor, that we call \emph{strain concentration tensor}, which depends on the normal vector $\vec n_\alpha$.
Here, we want to give some closed-form formulas for this tensor's coordinates, $A_{IJKL}$, in the cylinder basis $(\vec s_\alpha, \vec t_\alpha, \vec n_\alpha)$.

To that extent, we compute finite element (FE) solutions of Eshelby problem. We consider a spherical domain on the boundary of which we apply the uniform strain field $\tens{\overline{E}}$.
This spherical domain has a radius of $10L$. The mesh size in the cylinder is about $2R/3$.
For high contrasts and high aspect ratios, the significant values are $A_{nnnn}$, and $A_{ssnn}=A_{ttnn}$.
We computed these values for high contrasts $\chi=E_\alpha/E_0$ (between $10^2$ and $10^6$), different aspect ratios $e$ (between $40$ and $800$), and different values of Poisson's coefficients $\nu_0,\nu_\alpha$.
In fact, the value of $A_{nnnn}$ is practically the same whatever the value of fiber Poisson's coefficient $\nu_\alpha$, but, practically for all cases, we have $A_{ssnn}=-\nu_\alpha A_{nnnn}$. Moreover, when $\chi\geq10^5$, $\chi\,A_{nnnn}$ becomes independent of $\chi$. It suggests to write:
\begin{equation}
A_{nnnn}=A\,(e,\nu_0,\chi)/\chi
\end{equation}
where $A\,(e,\nu_0,\chi)$ becomes independent of $\chi$ when it is high. It also means that the values of $A\,(e,\nu_0,\chi)$ for every $e$, $\nu_0$ and $\chi$ are enough to know $A_{nnnn}$ and $A_{ssnn}$ (and also all coordinates $A_{IJKL}$ of $\tens A(\vec n_\alpha)$ in the cylinder basis).

\begin{figure}
\centering
\includegraphics[scale=1]{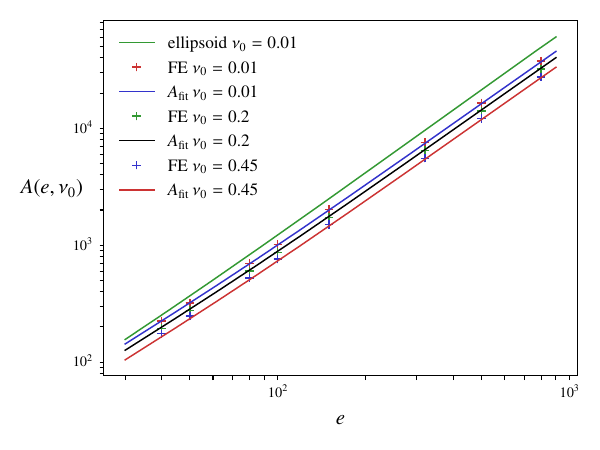}
\caption{$A\,(e,\nu_0,\chi)$, useful for computing the mean strain field solution of Eshelby's problem for a perfect cylinder. The FE values are computed here for a high contrast ($\chi=10^6$), for perfect cylinders ($e$ is the aspect ratio), different values of $\nu_0$ (Poisson's coefficient of the matrix), and for $\nu_\alpha=0.2$. '$A_{\mathrm{fit}}$' are the values obtained with function \eqref{A_fit_exp}. 'ellipsoid $\nu_0=0.01$' refers to the values $\chi\,A_{nnnn}$ for an ellipsoid, computed analytically with formulas found in \cite{torq2002}.}
\label{Annnn}
\end{figure}

We first present FE values of $A\,(e,\nu_0,\chi)$ for a contrast $\chi=10^6$, on Fig.~\ref{Annnn}. The aspect ratios vary between $40$ and $800$, Poisson's coefficients $\nu_0$ between $0.01$ and $0.45$, and we set $\nu_\alpha=0.2$.
We also choose to fit these FE values with a shape function,
inspired from expressions given in previous Section for a prolate ellipsoid ($e$ is the aspect ratio):
\begin{equation}
A_{\mathrm{fit}}\,(e,\nu_0)=(c+d\,\nu_0)\,e^a+(g+h\,\nu_0)(\ln e)^b
\end{equation}
where $a,b,c,d,g,h\in\reals$. We perform this fitting with function
\texttt{curve\_fit} of library \texttt{scipy} of \texttt{Python 3.8.3}.
We find $a=1.68$, $b=7.77$, $c=0.563$, $d=-0.340$, $g=-0.00194$, and $h=0.00115$:
\begin{equation}
\label{A_fit_exp}
A_{\mathrm{fit}}\,(e,\nu_0)=(0.563-0.340\,\nu_0)\,e^{1.68}+(-0.00194+0.00115\,\nu_0)(\ln e)^{7.77}
\end{equation}

We see on Fig.~\ref{Annnn} a very good agreement with the FE values.
Interestingly, we also computed the exact values of $\chi\,A_{nnnn}$ for an ellipsoid, for $\nu_0=0.01$, $\nu_\alpha=0.2$ and $\chi=10^6$, with the formulas given in \cite{torq2002} (without making any approximation of infinite contrast or high aspect ratio). We can see that ellipsoid values overestimate cylinder values, especially for very high aspect ratios.

We also computed the values of $A\,(e,\nu_0,\chi)$ for $\chi$ between $10^2$ and $10^6$ and we give these values in Table~\ref{table} (in Appendix). However, we were not able to find a good shape function which fits all the values.

\section{Average strain field on a distribution of fibers}
Let us now consider a distribution of fibers embedded in a uniform matrix.
The average strain field on all fibers $\langle\tens\varepsilon_\alpha\rangle$ can be computed by
\begin{equation}
\langle\tens\varepsilon_\alpha\rangle=\tens A_\alpha\dbldot\tens{\overline{E}}
\end{equation}
where $\tens A_\alpha$ is a fourth-rank tensor that we call \emph{mean strain concentration tensor}. We now
show how to compute this tensor.

\subsection{Change of basis}
We first need to transform the coordinates from the cylinder basis to the global basis.
We introduce the following matrix
\begin{equation}
\tens R=\begin{pmatrix}
\cos\theta\cos\phi&-\sin\phi&\sin\theta\cos\phi\\
\cos\theta\sin\phi&\cos\phi&\sin\theta\sin\phi\\
-\sin\theta&0&\cos\theta
\end{pmatrix}=\begin{pmatrix}
R_{1s}&R_{1t}&R_{1n}\\
R_{2s}&R_{2t}&R_{2n}\\
R_{3s}&R_{3t}&R_{3n}\\
\end{pmatrix}
\end{equation}
so that a vector $\vec u$ whose coordinates in the cylinder basis $(\vec s_\alpha, \vec t_\alpha, \vec n_\alpha)$ are $u_{I}$ ($I=s,t,n$), has the following coordinates in the global basis (using Einstein's notation):
\begin{equation}
u_i=R_{iI}u_I \quad (i=1,2,3).
\end{equation}

Moreover, considering a second-rank tensor field $\tens E$ whose coordinates in the cylinder basis $(\vec s_\alpha, \vec t_\alpha, \vec n_\alpha)$ are $E_{IJ}$ ($I,J=s,t,n$), its coordinates in the global basis will be
\begin{equation}
E_{ij}=R_{iI}R_{jJ}E_{IJ}\quad (i,j=1,2,3)\ \ \text{and}\ (I,J=s,t,n).
\end{equation}

Finally, considering a fourth-rank tensor field $\tens A$ whose coordinates in the cylinder basis $(\vec s_\alpha, \vec t_\alpha, \vec n_\alpha)$ are $A_{IJKL}$ ($I,J,K,L=s,t,n$),
its coordinates in the global basis will be
\begin{equation}
\label{change}
A_{ijkl}=R_{iI}R_{jJ}R_{kK}R_{lL}A_{IJKL}\quad (I,J,K,L=s,t,n)\ \ \text{and}\ \ (i,j,k,l=1,2,3).
\end{equation}
Now if we know the coordinates $A_{IJKL}$ of $\tens A(\vec n_\alpha)$ in the cylinder basis $(\vec s_\alpha, \vec t_\alpha, \vec n_\alpha)$,
its coordinates $A_{ijkl}$ in the global basis will be simply given by Eq.~\ref{change}.

\subsection{Isotropic distribution of fibers}
Let us consider a uniform isotropic distribution of fibers.
The coordinates of the mean strain concentration tensor $\tens A_\alpha$ in the global basis are
\begin{equation}
A_{ijkl}=\frac 1{4\pi}\int_{\phi=0}^{2\pi}\int_{\theta=0}^{\pi}U_{iIjJkKlL}(\theta,\phi)\sin\theta\,\D \theta\,\D \phi\,A_{IJKL}
\end{equation}
where $U_{iIjJkKlL}(\theta,\phi)=R_{iI}R_{jJ}R_{kK}R_{lL}$, and $A_{IJKL}$ are the coordinates in the fiber basis of the strain concentration tensors, introduced in Eq.~\ref{ellipsoid} or Eq.~\ref{cylinder}.

As we saw in Sec.~\ref{sec_closed} for high aspect ratios and high contrasts, only few components are significant in the fiber basis: $A_{nnnn}$, and $A_{ssnn}=A_{ttnn}$. We then have:
\begin{equation}
\begin{aligned}
A_{1111}&=\frac 2{\pi}W_5W_4A_{nnnn}+\frac 2{\pi}W_4(W_3-W_5)A_{ssnn}+\frac 2{\pi}W_3(W_2-W_4)A_{ttnn}\\
&=\frac 15A_{nnnn}+\frac{1}{20}A_{ssnn}+\frac{1}{12}A_{ttnn}\\
&=\frac 15A_{nnnn}+\frac{2}{15}A_{ssnn}
\end{aligned}
\end{equation}
where $W_n=\int_0^{\frac{\pi}{2}}\cos^n\phi\,\D\,\phi=\int_0^{\frac{\pi}{2}}\sin^n\phi\,\D\phi$ is Wallis integral.
In the same way, we compute the other terms:
\begin{equation}
\begin{aligned}
&A_{2222}=A_{3333}=A_{1111}\\
&A_{1122}=A_{2211}=A_{1133}=A_{3311}=A_{2233}=A_{3322}=\frac{1}{15}A_{nnnn}+\frac{4}{15}A_{ssnn}\\
&A_{1212}=A_{1313}=A_{2323}=\frac{1}{15}A_{nnnn}-\frac{1}{15}A_{ssnn}
\end{aligned}
\end{equation}
The other components are null.

\subsection{Planar distribution of fibers}
Let us consider a planar uniform distribution of fibers: fiber axes are all in the plane $(\vec e_1,\vec e_2)$. The coordinates of $\tens A_\alpha$ in the global basis are
\begin{equation}
A_{ijkl}=\frac 1{2\pi}\int_{\phi=0}^{2\pi}U_{iIjJkKlL}(\theta=\pi/2,\phi)\,\D \phi\,A_{IJKL}
\end{equation}
Considering again that $A_{nnnn}$, $A_{ssnn}$ and $A_{ttnn}$ are the only significant values, we have:
\begin{equation}
\begin{aligned}
A_{1111}&=A_{2222}=\frac{3}{8}A_{nnnn}+\frac{1}{8}A_{ssnn}\\
A_{3333}&=0\\
A_{1122}&=A_{2211}=\frac18A_{nnnn}+\frac38A_{ssnn}\\
A_{1133}&=A_{2233}=0\\
A_{3311}&=A_{3322}=\frac12A_{ssnn}\\
A_{1212}&=\frac18A_{nnnn}-\frac18A_{ssnn}\\
A_{2323}&=A_{1313}=0\\
\end{aligned}
\end{equation}
The other components are null.

\subsection{Unique orientation}
For cylinders oriented in the same direction, for example $\vec e_1$, the only significant components are:
\begin{equation}
A_{1111}=A_{nnnn},\quad A_{2211}=A_{3311}=A_{ssnn}.
\end{equation}

\section{Computation of some homogenization schemes}
The computation of homogenization schemes is straightforward. With previous results, we can compute
the mean strain concentration tensor $\tens A_\alpha$. With the formulas given below,
we then obtain the homogenized stiffness $\tens{C}^{\eff}$.

\subsection{Dilute scheme}
Dilute scheme gives:
\begin{equation}
\tens{C}^{\eff}=\tens{C}_0+f(\tens{C}_\alpha-\tens{C}_0)\dbldot\tens{A}_\alpha.
\end{equation}
Considering a high contrast, we have:
\begin{equation}
C^{\eff}_{ij}=C^0_{ij}+f\,C^\alpha_{ik}\,A^\alpha_{kj}.
\end{equation}
\paragraph{Isotropic distribution of fibers}
For an isotropic distribution of fibers:
\begin{equation}
\begin{aligned}
C^{\eff}_{11}&=C^0_{11}+f\,C^\alpha_{1k}\, A^\alpha_{k1}\\
&=C^0_{11}+f\left(2\nu_\alpha A^\alpha_{21}+(1-\nu_\alpha)A^{\alpha}_{11}\right)\frac{E_\alpha}{(1+\nu_\alpha)(1-2\nu_\alpha)}\\
&=C^0_{11}+f\left(\frac{3-\nu_\alpha}{15}A_{nnnn}+\frac{2+6\nu_\alpha}{15}A_{ssnn}\right)\frac{E_\alpha}{(1+\nu_\alpha)(1-2\nu_\alpha)}\\
\end{aligned}
\end{equation}
and
\begin{equation}
\begin{aligned}
C^{\eff}_{12}&=C^0_{12}+f\,C^\alpha_{1k}\, A^\alpha_{k2}\\
&=C^0_{12}+f\left((1-\nu_\alpha)A^{\alpha}_{12}+\nu_\alpha (A^\alpha_{22}+A^\alpha_{32})\right)\frac{E_\alpha}{(1+\nu_\alpha)(1-2\nu_\alpha)}\\
&=C^0_{12}+f\left(\frac{1+3\nu_\alpha}{15}A_{nnnn}+\frac{4+2\nu_\alpha}{15}A_{ssnn}\right)\frac{E_\alpha}{(1+\nu_\alpha)(1-2\nu_\alpha)}\\
\end{aligned}
\end{equation}
And if we now assume that $A_{ssnn}=-\nu_\alpha\,A_{nnnn}$,
\begin{equation}
\begin{aligned}
C^{\eff}_{11}&=C^0_{11}+f\frac{E_\alpha}{5}A_{nnnn}\\
C^{\eff}_{12}&=C^0_{12}+f\frac{E_\alpha}{15}A_{nnnn}.
\end{aligned}
\end{equation}
Because $\tens C^{\eff}$ is isotropic, these two components are enough to compute $E^\eff$ and $\nu^\eff$, using the following formulas:
\begin{equation}
\nu^\eff=\frac{C^\eff_{12}}{C^\eff_{12}+C^\eff_{11}},\qquad E^\eff=(1+\nu^\eff)\left(C^\eff_{11}-C^\eff_{12}\right)
\end{equation}

\paragraph{Planar distribution of fibers}
For a planar distribution, assuming $A_{ssnn}=-\nu_\alpha\,A_{nnnn}$,
\begin{equation}
\begin{aligned}
C^{\eff}_{11}&=C^0_{11}+f\,C^\alpha_{1k}\, A^\alpha_{k1}\\
&=C^0_{11}+f\left((1-\nu_\alpha)A^{\alpha}_{11}+\nu_\alpha A^\alpha_{21}+\nu_\alpha A^\alpha_{31}\right)\frac{E_\alpha}{(1+\nu_\alpha)(1-2\nu_\alpha)}\\
&=C^0_{11}+f\left(\frac{3-2\nu_\alpha}{8}A_{nnnn}+\frac{1+6\nu_\alpha}{8}A_{ssnn}\right)\frac{E_\alpha}{(1+\nu_\alpha)(1-2\nu_\alpha)}\\
&=C^0_{11}+f\frac{3-3\nu_\alpha-6\nu_\alpha^2}{8}A_{nnnn}\frac{E_\alpha}{(1+\nu_\alpha)(1-2\nu_\alpha)}\\
&=C^0_{11}+f\frac{3E_\alpha}{8}A_{nnnn}\\
\end{aligned}
\end{equation}
and of course, $C^{\eff}_{11}=C^{\eff}_{22}$. Moreover,
\begin{equation}
\begin{aligned}
C^{\eff}_{12}&=C^0_{12}+f\,C^\alpha_{1k}\, A^\alpha_{k2}\\
&=C^0_{12}+f\left((1-\nu_\alpha)A^{\alpha}_{12}+\nu_\alpha (A^\alpha_{22}+A^\alpha_{32})\right)\frac{E_\alpha}{(1+\nu_\alpha)(1-2\nu_\alpha)}\\
&=C^0_{12}+f\left(\frac{1+2\nu_\alpha}{8}A_{nnnn}+\frac{3+2\nu_\alpha}{8}A_{ssnn}\right)\frac{E_\alpha}{(1+\nu_\alpha)(1-2\nu_\alpha)}\\
&=C^0_{12}+f\frac{1-\nu_\alpha-2\nu_\alpha^2}{8}A_{nnnn}\frac{E_\alpha}{(1+\nu_\alpha)(1-2\nu_\alpha)}\\
&=C^0_{12}+f\frac{E_\alpha}{8}A_{nnnn}
\end{aligned}
\end{equation}
and of course, $C^{\eff}_{12}=C^{\eff}_{21}$. Besides,
\begin{equation}
\begin{aligned}
C^{\eff}_{31}&=C^0_{31}+f\,C^\alpha_{3k}\, A^\alpha_{k1}\\
&=C^0_{31}+f\left((1-\nu_\alpha)A^{\alpha}_{31}+\nu_\alpha (A^\alpha_{11}+A^\alpha_{21})\right)\frac{E_\alpha}{(1+\nu_\alpha)(1-2\nu_\alpha)}\\
&=C^0_{31}+f\left(\frac{\nu_\alpha}{2}A_{nnnn}+\frac{1}{2}A_{ssnn}\right)\frac{E_\alpha}{(1+\nu_\alpha)(1-2\nu_\alpha)}\\
&=C^0_{31}
\end{aligned}
\end{equation}
and of course, $C^{\eff}_{31}=C^{\eff}_{32}=C^{\eff}_{13}=C^{\eff}_{23}$. Moreover, $C^{\eff}_{33}=C^0_{33}$, $C^{\eff}_{44}=C^0_{44}$ and $C^{\eff}_{55}=C^0_{55}$ because $A^\alpha_{k3}=A^\alpha_{k4}=A^\alpha_{k5}=0$ for all k. Finally,
\begin{equation}
\begin{aligned}
C^{\eff}_{66}&=C^0_{66}+f\,C^\alpha_{66}\, A^\alpha_{66}\\
&=C^0_{66}+f\,\left(\frac14A_{nnnn}-\frac14A_{ssnn}\right)\frac{E_\alpha}{1+\nu_\alpha}\\
&=C^0_{66}+f\,\frac{E_\alpha}{4}A_{nnnn}\\
\end{aligned}
\end{equation}
which could be retrieved by the fact that $\tens C^\eff$ is the stiffness of a transverse isotropic material, then $C^\eff_{11}-C^\eff_{12}=C^\eff_{66}$.

\paragraph{Unique orientation of fibers}
For a unique orientation of fibers, we have $A^\alpha_{kj}=0$ if and only if $j\neq1$ or $k\geq 4$. Hence, assuming that $A_{ssnn}=-\nu_\alpha\,A_{nnnn}$,
\begin{equation}
\begin{aligned}
C^{\eff}_{11}&=C^0_{11}+f\,C^\alpha_{1k}\,A^\alpha_{k1}\\
&=C^0_{11}+f\,\left((1-\nu_\alpha)\,A_{nnnn}+2\nu_\alpha\,A_{ssnn}\right)\frac{E_\alpha}{(1+\nu_\alpha)(1-2\nu_\alpha)}\\
&=C^0_{11}+f\,A_{nnnn}\,E_\alpha
\end{aligned}
\end{equation}
And
\begin{equation}
\begin{aligned}
C^{\eff}_{21}&=C^0_{21}+f\,C^\alpha_{2k}\,A^\alpha_{k1}\\
&=C^0_{21}+f\,\left(\nu_\alpha A_{nnnn}+A_{ssnn}\right)\frac{E_\alpha}{(1+\nu_\alpha)(1-2\nu_\alpha)}\\
&=C^0_{21}
\end{aligned}
\end{equation}
and in the same way we show that $C^{\eff}_{ij}=C^0_{ij}$ for all $(i,j)\neq(1,1)$.

\subsection{Ponte-Casta\~neda \& Willis scheme for a uniform isotropic distribution of fibers}
Ponte-Casta\~neda \& Willis scheme with a uniform isotropic distribution of fibers gives the following effective stiffness:
\begin{equation}
\tens{C}^{\eff}=\tens{C}_0+f\left[\tens{Id}-f(\tens C_\alpha-\tens C_0)\dbldot\tens A_\alpha\dbldot\tens{P}_0\right]^{-1}\dbldot(\tens C_\alpha-\tens C_0)\dbldot\tens A_\alpha
\end{equation}
where $f$ is the volume fraction of fibers, and $\tens P_0$ is the Hill tensor of a sphere:
\begin{equation}
\tens P_0=\dfrac{1-2\nu_0}{6\mu_0(1-\nu_0)}\tens J+\dfrac{4-5\nu_0}{15\mu_0(1-\nu_0)}\tens K
\end{equation}
We can also write:
\begin{equation}
\tens{C}^{\eff}=\tens{C}_0+f\left[\tens A_\alpha^{-1}\dbldot(\tens C_\alpha-\tens C_0)^{-1}-f\tens{P}_0\right]^{-1}
\end{equation}
and for high contrasts:
\begin{equation}
\tens{C}^{\eff}=\tens{C}_0+f\left[\left(\tens C_\alpha\dbldot\tens A_\alpha\right)^{-1}-f\tens{P}_0\right]^{-1}.
\end{equation}
We already know, for an isotropic distribution of fibers, assuming $A_{ssnn}=-\nu_\alpha\,A_{nnnn}$, that
\begin{equation}
\begin{aligned}
C^\alpha_{1k}\,A^\alpha_{k1}&=\frac{E_\alpha}{5}A_{nnnn}=C^\alpha_{2k}\,A^\alpha_{k2}=C^\alpha_{3k}\,A^\alpha_{k3}\\
C^\alpha_{1k}\,A^\alpha_{k2}&=\frac{E_\alpha}{15}A_{nnnn}=C^\alpha_{1k}\,A^\alpha_{k3}=C^\alpha_{2k}\,A^\alpha_{k3}=C^\alpha_{2k}\,A^\alpha_{k1}=C^\alpha_{3k}\,A^\alpha_{k1}=C^\alpha_{3k}\,A^\alpha_{k2}.
\end{aligned}
\end{equation}
These components of $\tens C_\alpha\dbldot\tens A_\alpha$ are enough to compute the components
$C^\eff_{11}$ and $C^\eff_{12}$, which are also enough to compute $E^\eff$ and $\nu^\eff$.

\subsection{Mori-Tanaka scheme}
Mori-Tanaka scheme gives:
\begin{equation}
\tens{C}^{\eff}=\tens{C}_0+f(\tens{C}_\alpha-\tens{C}_0)\dbldot\tens{A}_\alpha\dbldot\left[f\tens A_\alpha+(1-f)\tens I\right]^{-1},
\end{equation}
and for high contrasts:
\begin{equation}
\tens{C}^{\eff}=\tens{C}_0+f\tens{C}_\alpha\dbldot\tens{A}_\alpha\dbldot\left[f\tens A_\alpha+(1-f)\tens I\right]^{-1}.
\end{equation}
Here again, the components of $\tens{C}_\alpha\dbldot\tens{A}_\alpha$ were given previously for the isotropic distribution, the planar distribution and the case of a unique orientation. It permits to compute easily the effective stiffness using Voigt notation.

\section*{Final remark}
If you notice an error in this document, do not hesitate to tell the writer at this adress:\\
antoin.martin@laposte.net\\
Besides, the writer will be very grateful for any other comment.

\section*{Appendix}
Fourth-rank tensors can be noted in a matrix form (Voigt notation)
\begin{equation}
\begin{aligned}
\tens A&=
\begin{pmatrix}
A_{1111}&A_{1122}&A_{1133}&\sqrt{2}A_{1123}&\sqrt{2}A_{1113}&\sqrt{2}A_{1112}\\
A_{2211}&A_{2222}&A_{2233}&\sqrt{2}A_{2223}&\sqrt{2}A_{2213}&\sqrt{2}A_{2212}\\
A_{3311}&A_{3322}&A_{3333}&\sqrt{2}A_{3323}&\sqrt{2}A_{3313}&\sqrt{2}A_{3312}\\
\sqrt{2}A_{2311}&\sqrt{2}A_{2322}&\sqrt{2}A_{2333}&2A_{2323}&2A_{2313}&2A_{2312}\\
\sqrt{2}A_{1311}&\sqrt{2}A_{1322}&\sqrt{2}A_{1333}&2A_{1323}&2A_{1313}&2A_{1312}\\
\sqrt{2}A_{1211}&\sqrt{2}A_{1222}&\sqrt{2}A_{1233}&2A_{1223}&2A_{1213}&2A_{1212}\\
\end{pmatrix}\\
&=\begin{pmatrix}
A_{11}&A_{12}&A_{13}&A_{14}&A_{15}&A_{16}\\
A_{21}&A_{22}&A_{23}&A_{24}&A_{25}&A_{26}\\
A_{31}&A_{32}&A_{33}&A_{34}&A_{35}&A_{36}\\
A_{41}&A_{42}&A_{43}&A_{44}&A_{45}&A_{46}\\
A_{51}&A_{52}&A_{53}&A_{54}&A_{55}&A_{56}\\
A_{61}&A_{62}&A_{63}&A_{64}&A_{65}&A_{66}\\
\end{pmatrix}
\end{aligned}
\end{equation}
when they respect 'minor symmetries': $A_{ijkl}=A_{jikl}=A_{ijlk}$. Note that the 'major symmetry' may not be true ($A_{ijkl}\neq A_{klij}$).
For the tensors $\tens I$, $\tens J$ and $\tens K$ introduced above, it gives:
\begin{equation}
\begin{aligned}
&\tens I=
\begin{pmatrix}
1&0&0&0&0&0\\
0&1&0&0&0&0\\
0&0&1&0&0&0\\
0&0&0&1&0&0\\
0&0&0&0&1&0\\
0&0&0&0&0&1\\
\end{pmatrix}\quad \tens J=\begin{pmatrix}
1/3&1/3&1/3&0&0&0\\
1/3&1/3&1/3&0&0&0\\
1/3&1/3&1/3&0&0&0\\
0&0&0&0&0&0\\
0&0&0&0&0&0\\
0&0&0&0&0&0\\
\end{pmatrix}\\
&\\
&\tens K=\begin{pmatrix}
2/3&-1/3&-1/3&0&0&0\\
-1/3&2/3&-1/3&0&0&0\\
-1/3&-1/3&2/3&0&0&0\\
0&0&0&1&0&0\\
0&0&0&0&1&0\\
0&0&0&0&0&1\\
\end{pmatrix}
\end{aligned}
\end{equation}
This notation allows to perform the tensor double-contraction $\tens A\dbldot\tens B$ as a standard $6\times6$ - matrix product, and the tensor inversion as a standard matrix inversion.

\begin{table}
\centering
  \begin{tabular}{ c }
$\chi=10^2$
\end{tabular}\\
  \begin{tabular}{| c | c | c | c | c | c | c | c |}
\hline
$e$&$\nu_0=0.01$&$\nu_0=0.05$&$\nu_0=0.1$&$\nu_0=0.2$&$\nu_0=0.3$&$\nu_0=0.4$&$\nu_0=0.45$\\
\hline
$40$&$65.0$&$64.4$&$63.6$&$62.2$&$61.1$&$60.5$&$60.7$\\
\hline
$50$&$71.2$&$70.7$&$70.0$&$68.8$&$67.8$&$67.2$&$67.3$\\
\hline
$80$&$81.5$&$81.1$&$80.6$&$79.8$&$79.1$&$78.7$&$78.7$\\
\hline
$100$&$85.1$&$84.8$&$84.4$&$83.7$&$83.1$&$82.8$&$82.8$\\
\hline
$150$&$90.0$&$89.8$&$89.5$&$89.0$&$88.7$&$88.4$&$88.4$\\
\hline
$320$&$95.3$&$95.2$&$95.0$&$94.8$&$94.6$&$94.5$&$94.5$\\
\hline
$500$&$97.0$&$96.9$&$96.8$&$96.7$&$96.6$&$96.5$&$96.5$\\
\hline
$800$&$98.1$&$98.1$&$98.0$&$97.9$&$97.8$&$97.8$&$97.8$\\
\hline
\end{tabular}
\hspace{2cm}
 \begin{tabular}{ c }
\\
\end{tabular}\\
  \begin{tabular}{ c }
$\chi=10^3$
\end{tabular}\\
  \begin{tabular}{| c | c | c | c | c | c | c | c |}
\hline
$e$&$\nu_0=0.01$&$\nu_0=0.05$&$\nu_0=0.1$&$\nu_0=0.2$&$\nu_0=0.3$&$\nu_0=0.4$&$\nu_0=0.45$\\
\hline
$40$&$180.0$&$175.1$&$169.6$&$160.1$&$152.7$&$147.7$&$146.8$\\
\hline
$50$&$236.5$&$230.6$&$223.7$&$211.9$&$202.4$&$195.7$&$194.2$\\
\hline
$80$&$390.0$&$382.4$&$373.4$&$357.5$&$344.3$&$334.1$&$331.0$\\
\hline
$100$&$472.5$&$464.6$&$455.4$&$438.8$&$424.6$&$413.5$&$409.8$\\
\hline
$150$&$617.2$&$610.2$&$601.9$&$586.6$&$573.2$&$562.3$&$558.4$\\
\hline
$320$&$811.7$&$807.8$&$803.2$&$794.4$&$786.6$&$780.0$&$777.4$\\
\hline
$500$&$878.6$&$876.1$&$873.0$&$867.3$&$862.2$&$857.8$&$856.1$\\
\hline
$800$&$923.8$&$922.2$&$920.3$&$916.7$&$913.4$&$910.7$&$909.6$\\
\hline
\end{tabular}
\hspace{2cm}
\begin{tabular}{ c }
\\
\end{tabular}\\
  \begin{tabular}{ c }
$\chi=10^4$
\end{tabular}\\
  \begin{tabular}{| c | c | c | c | c | c | c | c |}
\hline
$e$&$\nu_0=0.01$&$\nu_0=0.05$&$\nu_0=0.1$&$\nu_0=0.2$&$\nu_0=0.3$&$\nu_0=0.4$&$\nu_0=0.45$\\
\hline
$40$&$219.7$&$212.5$&$204.4$&$190.8$&$180.2$&$173.1$&$171.7$\\
\hline
$50$&$310.7$&$300.5$&$289.0$&$269.5$&$254.2$&$243.5$&$240.7$\\
\hline
$80$&$645.9$&$625.2$&$601.6$&$561.3$&$529.0$&$504.7$&$496.8$\\
\hline
$100$&$908.5$&$880.2$&$847.7$&$791.8$&$746.6$&$712.0$&$699.8$\\
\hline
$150$&$1642.1$&$1595.0$&$1540.7$&$1445.9$&$1367.7$&$1305.8$&$1282.3$\\
\hline
$320$&$4054.3$&$3975.4$&$3882.5$&$3714.0$&$3568.0$&$3445.5$&$3395.0$\\
\hline
$500$&$5752.3$&$5678.2$&$5589.4$&$5424.9$&$5277.8$&$5150.4$&$5096.2$\\
\hline
$800$&$7231.5$&$7177.0$&$7111.1$&$6987.1$&$6873.9$&$6773.5$&$6729.9$\\
\hline
\end{tabular}
\hspace{2cm}
\begin{tabular}{ c }
\\
\end{tabular}\\
  \begin{tabular}{ c }
$\chi=10^5$
\end{tabular}\\
  \begin{tabular}{| c | c | c | c | c | c | c | c |}
\hline
$e$&$\nu_0=0.01$&$\nu_0=0.05$&$\nu_0=0.1$&$\nu_0=0.2$&$\nu_0=0.3$&$\nu_0=0.4$&$\nu_0=0.45$\\
\hline
$40$&$224.7$&$217.2$&$208.7$&$194.5$&$183.5$&$176.1$&$174.7$\\
\hline
$50$&$320.8$&$310.0$&$297.7$&$277.0$&$260.9$&$249.6$&$246.6$\\
\hline
$80$&$691.7$&$668.0$&$641.1$&$595.5$&$559.2$&$532.1$&$523.2$\\
\hline
$100$&$1002.2$&$967.8$&$928.7$&$861.9$&$808.6$&$767.9$&$753.6$\\
\hline
$150$&$1977.8$&$1909.8$&$1832.5$&$1699.8$&$1592.5$&$1508.8$&$1477.1$\\
\hline
$320$&$6984.3$&$6754.7$&$6491.7$&$6035.1$&$5658.9$&$5355.4$&$5232.6$\\
\hline
$500$&$14068.3$&$13643.4$&$13152.8$&$12291.7$&$11570.9$&$10978.4$&$10732.8$\\
\hline
$800$&$26923.9$&$26247.8$&$25457.6$&$24043.7$&$22831.2$&$21809.3$&$21375.1$\\
\hline
\end{tabular}
\hspace{2cm}
\begin{tabular}{ c }
\\
\end{tabular}\\
\end{table}

\begin{table}
\centering
  \begin{tabular}{ c }
$\chi=10^6$
\end{tabular}\\
  \begin{tabular}{| c | c | c | c | c | c | c | c |}
\hline
$e$&$\nu_0=0.01$&$\nu_0=0.05$&$\nu_0=0.1$&$\nu_0=0.2$&$\nu_0=0.3$&$\nu_0=0.4$&$\nu_0=0.45$\\
\hline
$40$&$225.2$&$217.7$&$209.2$&$194.8$&$183.9$&$176.4$&$175.0$\\
\hline
$50$&$321.8$&$310.9$&$298.6$&$277.8$&$261.6$&$250.2$&$247.2$\\
\hline
$80$&$696.6$&$672.6$&$645.4$&$599.1$&$562.4$&$535.0$&$526.0$\\
\hline
$100$&$1012.6$&$977.5$&$937.6$&$869.6$&$815.3$&$774.0$&$759.5$\\
\hline
$150$&$2019.2$&$1948.4$&$1868.0$&$1730.3$&$1619.2$&$1532.7$&$1499.9$\\
\hline
$320$&$7534.1$&$7267.6$&$6963.9$&$6441.1$&$6014.2$&$5672.2$&$5534.4$\\
\hline
$500$&$16499.6$&$15918.1$&$15254.3$&$14107.7$&$13165.9$&$12403.2$&$12089.8$\\
\hline
$800$&$37498.1$&$36201.2$&$34716.6$&$32141.1$&$30010.5$&$28267.4$&$27540.0$\\
\hline
\end{tabular}
\caption{Useful values $A\,(e,\nu_0,\chi)$ for computing the mean strain field solution of Eshelby's problem for a perfect cylinder. The values are computed for different contrasts $\chi$, for perfect cylinders ($e$ is the aspect ratio), different values of $\nu_0$ (Poisson's coefficient of the matrix), and for $\nu_\alpha=0.2$.}
\label{table}
\end{table}

\end{document}